\documentclass[useAMS,usenatbib,usegraphicx]{mn2e}

\usepackage{color}

\renewcommand{\mathbf}{\bmath}
\renewcommand{\vec}{\mathbf}
\newcommand{\eq}[1]{Eq.\ (\ref{#1})}

\def\be{\begin{equation}}
\def\ee{\end{equation}}
\def\baray{\begin{eqnarray}}
\def\ba{\baray}
\def\earay{\end{eqnarray}}
\def\ea{\earay}
\def\bal#1\eal{\begin{align}#1\end{align}} 

\def\Mev{{\rm MeV}}
\def\nnuc{n_{\rm nuc}}
\def\mpstar{m_p^\star}
\def\tcp{T_{c,p}}
\def\tcpn{T_{c,p,9}}
\def\msun{M_\odot}

\def\grad{{\mbox{\boldmath $\nabla$}}}
\def\dotprod{{\mbox{\boldmath $\cdot$}}}
\def\crossprod{{\mbox{\boldmath $\times$}}}

\def\curl{\grad\crossprod}
\def\Bvec{{\mbox{\boldmath $B$}}}
\def\Bhat{{\mbox{\boldmath ${\hat B}$}}}
\def\nhat{{\mbox{\boldmath ${\hat n}$}}}

\def\rhat{{\mbox{\boldmath ${\hat r}$}}}
\def\zhat{{\mbox{\boldmath ${\hat z}$}}}
\def\phihat{{\mbox{\boldmath ${\hat\phi}$}}}
\def\thetahat{{\mbox{\boldmath ${\hat\theta}$}}}
\def\Hvec{{\mbox{\boldmath $H$}}}

\def\epsnuc{\epsilon_{\rm nuc}}
\def\munuc{\mu_{\rm nuc}}

\def\mmax{M_{\rm max}}

\def\Jcal{{\cal J}}
\def\Jch{{\hat\Jcal}}
\def\Jhat{{\hat\Jcal}}

\def\ftil{{\tilde f}}
\def\dA{\Delta A}
\def\Scal{{\cal S}}
\def\jmax{{j_{\rm max}}}
\def\pfp{p_{F,p}}
\def\epsb{\epsilon_b}
\def\Jchshell{\Jch_{\rm shell}}
\def\Jchcore{\Jch_{\rm core}}
\def\delb{\delta_b}
\def\Jchc{\Jch_{\rm II}}
\def\dJchc{\dJch_{\rm II}}
\def\dJchs{\dJch_{\rm shell}}
\def\pj{P_{2j+1}{}^1}
\def\dH{\Delta B}
\def\thm{{\theta_{\rm II}(x)}}
\def\dJchsn{\Delta\Jch_{n,{\rm shell}}}
\def\dJchcn{\Delta\Jch_{n,{\rm II}}}
\def\Dnj{{\cal{N}}_{n,j}(x)}

\def\Sz{S^{(0)}}
\def\Jchz{\Jch^{(0)}}
\def\qcore{q_{2,{\rm II}}}
\def\qshell{q_{2,{\rm shell}}}
\def\Qshell{Q_{2,{\rm shell}}}
\def\pj{P_{2j+1}{}^1}
\def\Mcal{{\cal M}}
\def\Ical{{\cal I}}
\def\dJch{\Delta\Jch}
\def\dIcal{\Delta\Ical}
\def\dIcalc{\dIcal_{\rm II}}
\def\dIcals{\dIcal_{\rm shell}}
\def\fHB{f\left(\sqrt{\frac{\Hc(s)}{B}}\right)}
\def\Hc{H_{c1}}

\def\xhat{{\mathbf{\hat x}}}
\def\yhat{{\mathbf{\hat y}}}
\def\zhat{{\mathbf{\hat z}}}
\def\gtrsim{\ga}
\def\lesssim{\la}
\def\onehalf{{\frac{1}{2}}}
\def\onethird{{\frac{1}{3}}}

\onecolumn

\title[Poloidal Magnetic Fields In Superconducting Neutron Stars]
{Poloidal Magnetic Fields In Superconducting Neutron Stars}
\author[K. T. Henriksson and I. Wasserman]
{K. T. Henriksson\thanks{{Email: kth48@cornell.edu}} and I. Wasserman\thanks{Email: ira@astro.cornell.edu}\thanks{Corresponding Author}\\
 Center for Radiophysics and Space Research, Cornell University, Ithaca, NY 14853, USA}

\begin{document}


\pagerange{\pageref{firstpage}--\pageref{lastpage}} \pubyear{2012}

\maketitle

\label{firstpage}
\begin{abstract}
We develop the formalism for computing the magnetic field within an axisymmetric neutron star with a
strong Type II superconductor core surrounded by a normal conductor. 
The formalism takes full account
of the constraints imposed by hydrostatic equilibrium with a barotropic equation of state.
A characteristic of this problem is that the currents and fields need to be determined
simultaneously and self-consistently. Within the core, the strong Type II limit $B\ll H$
allows us to compute the shapes of individual field lines. 
We specialize to purely poloidal magnetic fields
that are perpendicular to the equator, and develop the ``most dipolar case'' in which field
lines are vertical at the outer radius of the core, which leads to a
magnetic field at the stellar surface that is as close to a dipole as possible.
We demonstrate that although field lines from the core may only penetrate a short distance into
the normal shell, boundary conditions at the inner radius of the normal shell control the
field strength on the surface. Remarkably, we find that for a Newtonian $N=1$ polytrope,
the surface dipole field strength is $B_{\rm surf}\simeq H_b\epsb/3$ where $H_b$ is the 
magnetic field strength at the outer boundary of the Type II core and $\epsb R$ is the
thickness of the normal shell. For reasonable models, $H_b\approx 10^{14}$ G and $\epsb
\approx 0.1$ so the surface field strength is $B_{\rm surf}\simeq 3\times 10^{12}$ G,
comparable to the field strengths of many radiopulsars. In general, $H_b$ and $\epsb$
are both determined by the equation of state of nuclear matter and by the mass of the
neutron star, but $B_{\rm surf}\sim 10^{12}$ G is probably a robust result for the
``most dipolar'' case. We speculate on how the wide range of neutron star surface fields
might arise in situations with less restrictions on the internal field configuration.
We show that quadrupolar distortions are $\sim -10^{-9}(H_b/10^{14}\,{\rm G})^2$ and 
arise primarily in the normal shell for $B\ll H_b$.

\end{abstract}
\begin{keywords}
stars: neutron; magnetic fields
\end{keywords}

\section{Introduction}

Theoretical arguments predict that the protons in the
core of a neutron star form a superconductor
\citep{BAYM1969},
which will be Type II provided that $\kappa=1/k_L\xi_p
>1/\sqrt{2}$, where $\xi_p$ is the coherence length and
$1/k_L$ is the London length
\citep{PhysRevD.16.275};
numerically
{
\ba
k_L^2&=&{4\pi n_pe^2\over \mpstar c^2}={4\pfp^3e^2\over 3\pi\hbar^3\mpstar c^2}
=(306\,{\rm fm})^{-2}\left({\pfp\over 50\,{\rm
MeV}}\right)^3\left(\frac{m_p}{\mpstar}\right)
\nonumber\\
\xi_p&=&{\hbar\pfp\over \pi\Delta_p\mpstar}={3.33\,{\rm
fm}(\pfp/50\,{\rm MeV})\over\Delta_p({\rm MeV})(\mpstar/m_p)}
\nonumber\\
\kappa&=&{1\over k_L\xi_p}={91.8\Delta_p({\rm MeV})(\mpstar/m_p)^{3/2}\over
(\pfp/50\,{\rm MeV})^{5/2}}
\label{typeiicond}
\ea
where $\mpstar$ is the proton effective mass
\citep[e.g.][]{1975JETP...42..164A,1984ApJ...282..533A,1995ApJ...447..305S,2012arXiv1209.1776Z}.
The gap energy in MeV is}
$\Delta_p(\Mev)=0.15\tcpn$ for a critical temperature
$\tcp=10^9\tcpn\,{\rm K}$
(because the pairing is $^1S_0$) and $n_p=\pfp^3/3\pi^2\hbar^3$
is the proton density. We expect $\pfp\simeq 50$ MeV at
or just below nuclear density, $\nnuc=0.16\,{\rm fm}^{-3}$,
which is near the boundary between the superconducting
core and normal conducting crust inside a neutron star
\citep{Baym1971225,2001LNP...578..127H}
Although calculations of $\Delta_p$ are still being refined
\citep{PhysRevC.78.015805, Zuo200444}, \eq{typeiicond} is
likely to be consistent with Type II superconductivity for $n_b\lesssim 2\nnuc$.

Two recent observations are relevant to this question. First, 
X-ray observations of the Cassiopeia  A neutron star are consistent
with cooling via the Cooper pair formation process
\citep{1976ApJ...205..541F, MNL2:MNL21015, PhysRevLett.106.081101},
and require proton superconductivity to suppress the contribution
from URCA processes, particularly Modified URCA; if
$\tcpn\gtrsim 2-3$ throughout the Cassiopeia A neutron star, these 
processes are unimportant \citep{MNL2:MNL21015}. Second,
the discovery of a $2\msun$ neutron star favors relatively
stiff equations of state of nuclear matter, in which case
internal densities may not be far above $\nnuc$
\citep{Demorest2010}.
We can quantify using relativistic polytropes, where the relationship
between pressure $P$ and baryon density $n_b$ is $P\propto
n_b^{1+1/N}$; to fix the scale we use the
chemical potential is $\munuc=m_b(1+\epsnuc)$ at
nuclear density, where $m_b$ is the baryon mass.
Requiring (i) $\epsnuc=0.065$, a reasonable value
\citep{Hebeler2007176},
(ii) $\mmax>2\msun$ and
(iii) sound speed $<1$ implies $0.54\leq N\leq 0.73$,
for which $\mmax\approx 5.44(\epsnuc)^{N/2}\msun\simeq
2.0-2.6\msun$, and central densities are 
$n_b/\nnuc\simeq 1.7-2.4$ and $n_b/\nnuc\simeq 2.4-6.7$ 
for $M=1.4\msun$ and $M=2.0\msun$, respectively.
Actual equations
of state may be stiffer than $N=1$ in some density ranges,
and softer in others, but central densities may not be 
much larger than $\simeq 2\nnuc$ for ``typical'' neutron
stars. In this relatively limited density range, and with
$\tcpn\gtrsim 2-3$, $\kappa >1/\sqrt{2}$
is likely to be true, so the superconductor is Type II.

Inside a Type II superconductor, magnetic flux is concentrated
within flux tubes, each of which carries magnetic flux
$\Phi_0=\pi\hbar c/e\approx 2.06\times 10^{-7}\,{\rm G\, cm^2}$.
The spacing between flux tubes 
{ in a triangular lattice is 
$\ell_B=(2\Phi_0/B\sqrt{3})^{1/2}= 4880B_{12}^{-1/2}$ fm, where 
$B=10^{12}B_{12}$ G is the magnitude of the
magnetic induction in the core; consequently
\be
k_L\ell_B=15.9B_{12}^{-1/2}\left({\pfp\over 50\,{\rm MeV}}\right)^{3/2}
\left(\frac{m_p}{\mpstar}\right)^{1/2}~.
\label{strongtypeiicond}
\ee
When} $k_L\ell_B\gg 1$, the magnetic field is confined to flux tubes
that do not interact with one another to a first approximation; the
local magnetic field near a flux tube decays exponentially with a
scale length $1/k_L$. Under these circumstances, the magnetic free
energy is nearly proportional to $B$ and the magnetic field strength
inside the superconductor is approximately
\citep{PhysRevD.16.275}
\be
H\simeq\Hc\simeq{\Phi_0 k_L^2\ln\kappa\over 4\pi}=
{\pfp^3e\ln\kappa\over 3\pi\mpstar c\hbar^2}=
8.75\times 10^{13}{\rm G}\,\left({\ln\kappa\over 5}\right)\left({\pfp\over 50\,{\rm
MeV}}\right)^3
\left(\frac{m_p}{\mpstar}\right)
\label{Hval}
\ee
which is a function of density, since $\pfp$, $\mpstar$ and $\kappa$ vary with density according
to the neutron star equation of state;
from the calculations in \citet{PhysRevC.78.015805}, we conclude that a fairly good
approximation is $H\propto\rho^b$ with $b\approx1.6-1.8$.
Since $k_L\ell_B$ is large, the superconducting core of a neutron star  
is in the ``strong Type II''  regime. That the
magnetic field strength in this regime is fixed by the density both simplifies and 
complicates the computation of magnetic structure. The strong Type II limit fails
within a (presumably thin) transition region within which protons cluster progressively
into nuclei until the superconducting free proton fluid disappears entirely.
The outer shell of the neutron star is then a normal conductor.

The challenge is to compute the magnetic field structure including both Type II inner
core and normal outer shell, which then matches to the (vacuum) exterior. In this 
paper, we focus on the ``most dipolar'' external fields within stars we model as 
$N=1$ Newtonian polytropes. In order to find these most dipolar configurations, we require that
field lines exit the core vertically. We then show that if $H_b\gg B$ is the magnetic 
field strength at the outer edge of the Type II core, then 
the magnetic field at the 
base of the normal shell is very nearly $-\thetahat H_b\sin\theta$, provided that the
transition region between the core and normal shell is thin enough.
This boundary condition is consistent with a particular dipole field solution
within the normal shell. From this solution we find that the characteristic magnetic dipole
field strength at the stellar surface under these conditions is approximately
\be
{\mu\over R^3}\simeq {H_b\epsb\over 3}\simeq 2.91\times 10^{12}
{\rm G}\,\left({\ln\kappa\over 5}\right)\left({\pfp\over 50\,{\rm
MeV}}\right)^3\left(\frac{m_p}{\mpstar}\right)\left({\epsb\over 0.1}\right)~.
\label{dipolefieldstrength}
\ee
where the thickness of the crust is $\epsb R$ and $\mu$ is the dipole moment. 

Eq. (\ref{dipolefieldstrength})
is one of the central results of this paper: it relates the 
dipole magnetic field strength directly to the magnetic field in
the Type II core and the thickness of the stellar crust, both of
which may be computed given the nuclear equation of state. 
Although Eq. (\ref{dipolefieldstrength}) depends sensitively on
$\pfp$ at the boundary of the Type II core, it is noteworthy that
the implied field strength in this model is comparable to
$B\sim 10^{12}$ G, which is characteristic of many pulsars and
accreting neutron stars. Detailed results will differ among
equations of state; in particular, $\epsb\propto\rho_b^{1/N}$,
where $\rho_b$ is the density at the base of the normal shell,
for a polytrope of index $N$.

{Other calculations based on magneto-thermal effects and evolution have also
arrived at surface magnetic field strengths of order $10^{12}-10^{13}$ G
\citep{1983MNRAS.204.1025B,2009A&A...496..207P}. These arguments did not consider
the Type II core, and the significant boundary condition it imposes at
the base of the neutron star's normal outer shell.}

In \S\ref{overview} we present a general overview of the problem we
address here. In this section, we derive (i) the implications of
hydrostatic balance for magnetic field configurations, (ii) the
equations that we solve in the Type II core, (iii) the equations
that hold in the normal shell, and (iv) the formalism for computing
stellar distortions. We give specific results relevant to the ``most
dipolar'' case and derive, in particular, \eq{dipolefieldstrength},
but also highlight causes of perturbations around this simple
solution.
In \S\ref{toy} we illustrate some of the salient features
of the problem
via a toy model that is totally analytic and surprisingly close
to being realistic. 
Then in \S\ref{solutions} we present particular
results for the most dipolar case of greatest interest. In this 
section, we consider not only the simplest version of the ``most
dipolar'' case but also perturbations around that solution that lead
to non-dipolar corrections to the fields.  We conclude in \S\ref{conclusions}
by reviewing these solutions and suggesting extensions to other
cases that may allow surface field strengths that are either appreciably
lower or higher than Eq. (\ref{dipolefieldstrength}).

\section{Overview}
\label{overview}

In this paper, we compute the magnetic field for a star
with a superconducting core that matches onto a normal 
crust surrounded by vacuum. We model the star as 
a Newtonian $N=1$ polytrope with a poloidal magnetic field
that only distorts the star slightly.
Previous calculations of this
type were done for poloidal fields of a uniform density star 
\citep{ROBERTS01081981}
and for toroidal magnetic fields in a Newtonian
$N=1$ polytrope
{\citep{2008MNRAS.383.1551A, AkgunThesis, 2012MNRAS.419..732L}}.
These two different previous calculations were simpler 
than those presented here. A significant complication
in finding the magnetic field structure is that 
$H$ is density-dependent according to \eq{Hval};
accommodating this complication will require a generalization
of the method of solution for a uniform density star
\citep{ROBERTS01081981}.
Calculations for a toroidal field could account for the
density dependence of $H$ because it was unnecessary to
determine the field line shapes, which were specified
{\sl a priori} \citep{2008MNRAS.383.1551A}.

Our calculations are perturbative in that we assume that
the distortions induced by magnetic stresses (which we
compute) are small.
However, we take full account of the 
requirements of hydrostatic balance, which constrain the
field shapes, just as for normal conductors
\citep{1961ApJ...133..170W, 1956ApJ...123..498P}; our calculations are 
analogous to those done previously for small magnetic
distortions in a normal conductor \citep{1965MNRAS.131..105M}.
We find the ``true equilibrium'' configurations in which the 
magnetic distortions are assumed to obey the same equation of state
as the unperturbed star. The actual magnetic fields in neutron
stars may not be true equilibria in this sense
\citep{2011MNRAS.417.2288M}
even though, on sufficiently long time scales, the fields should
relax to these states naturally. Even in non-equilibrium field
configurations the field at the base of the normal shell will
have to match properly to the field in the Type II core. We 
therefore expect a version of Eq. (\ref{dipolefieldstrength})
to remain true for substantially dipolar fields even for
configurations that are not true equilibria.

We focus on the ``most dipolar'' field configurations that arise
when field lines hit the outer edge of the Type II core vertically.
As we shall see, this is a special configuration, and infinitely
many others are possible. 

Throughout this paper, we assume that the entire core of
the star is superconducting. We argued above that this may
be reasonable if the density range in the core is not 
too large. We expect that the external dipole field that emerges
from our ``most dipolar'' solutions would not be affected 
significantly if the very inner core of the neutron star is
not superconducting.

\subsection{Requirements of Hydrostatic Balance\label{sec:hydro}}

The magnetic free energy is $f(\rho, \vec{B})$, where $\rho$ is mass
density and $\vec{B}$ is magnetic induction; then the magnetic field 
is $\Hvec=4\pi\partial f/\partial\vec{B}$,
and the magnetic force density is
\be
\mathbf{f}^{mag}
=-\rho\grad\left(\frac{\partial f}{\partial\rho}\right)+\frac{(\curl\vec{H})\crossprod\vec{B}}{4\pi}
=-\rho\grad\left(\frac{\partial f}{\partial\rho}\right)+\frac{\vec{J}\crossprod\vec{B}}{c}
\label{magforce}
\ee
using Amp\'ere's law, $\curl\vec{H}=4\pi\vec{J}/c$~.
{\eq{magforce} was derived in \citet{2008MNRAS.383.1551A} by taking the divergence of the
magnetic stress tensor for a Type II superconductor given by \citet{PhysRevD.16.275}; it may also
be derived by considering the variation of magnetic energy resulting from small fluid displacements.}
The equation of hydrostatic balance including pressure, gravity and
magnetic forces is
\be
0=-\frac{\grad P}{\rho}-\grad\Psi+\frac{\mathbf{f}^{mag}}{\rho}
\ee
 where $\Psi$ is the gravitational potential, $P$ is the pressure
and $\rho$ is the mass density. For a barotropic fluid
$P=P(\rho)$ and $dh(\rho)=dP(\rho)/\rho$, in which case 
\be
0=-\grad\left(h+\Psi+\frac{\partial f}{\partial\rho}\right)+\frac{\vec{J}\crossprod\vec{B}}{\rho c}~.
\label{barohydro}
\ee
Eq. (\ref{barohydro}) is only consistent mathematically if $\vec{J}\crossprod\vec{B}/\rho c=-\grad
\Phi$, where $\Phi$ is some potential. In axisymmetry,
\be
\vec{B}=\curl\left[\frac{A(r,\theta)\phihat}{r\sin\theta}\right]+
B_T(r,\theta)\phihat
\label{axiB}
\ee
and
$\vec{H}=\vec{H}_P+H_T\phihat$; the current density is
\be
\curl\vec{H}=\curl\vec{H}_P+\frac{\grad(H_Tr\sin\theta)\crossprod\phihat}{r\sin\theta}
\equiv {4\pi J\phihat\over c}+\frac{\grad(H_Tr\sin\theta)\crossprod\phihat}{r\sin\theta}~.
\label{currentdensity}
\ee
Requiring that $\phihat\dotprod(\vec{J}\crossprod\vec{B})=0$ implies that 
$H_Tr\sin\theta=\mathcal{H}_T(A)$; the Lorentz acceleration
$\vec{J}\crossprod\vec{B}/\rho c$ is a total 
gradient if 
\be
J-\frac{cB_T}{4\pi}\frac{d{\mathcal H}_T(A)}{dA}=c\rho r\sin\theta\mathcal{J}(A)
\label{jcalA}
\ee 
{which means that $-\grad\Phi=\mathcal{J}(A)\grad A$.}
\eq{jcalA} is
well-known 
for normal magnetic equilibria
\citep{1956ApJ...123..498P, 1960ApJ...131..227W, 1961ApJ...133..170W, 1965MNRAS.131..105M};
the $B_T=0$ version of \eq{jcalA}
has also been derived previously for Type II superconductors with poloidal
fields
\citep{ROBERTS01081981,AkgunThesis}.
{Similar results have been derived independently by \citet{2013PhRvL.110g1101L}.}

We are interested in poloidal fields only, so $B_T=0$. 
{\citep[See also][for specific models including poloidal and toroidal fields.]{2013PhRvL.110g1101L}}
We also
assume that $f(\rho,\vec{B})=f(\rho,B)$ i.e. there are no preferred directions
in space so the free energy only depends on $B=\vert\vec{B}\vert$. In this
situation, the magnetic field and magnetic induction are parallel:
$\Hvec=4\pi\Bhat\partial f/\partial B$ where $\Bhat=\vec{B}/B$.

Our goal is to solve 
\be
\curl\Hvec=\curl(H\Bhat)=\grad H\crossprod\Bhat+H\curl\Bhat=
4\pi\rho r\sin\theta\Jcal(A)\phihat
\label{Ampere}
\ee 
throughout the star. We shall strive for solutions without surface currents;
in particular, we are interested in solutions in which field lines cross 
the equator smoothly and vertically. With this goal in mind, we will have
to determine {\sl both} the field configuration {\sl and} its source
$\Jcal(A)$ self consistently, within both the Type II core and the normal
shell. 

\subsection{Field Configuration in the Type II Core}
\label{TypeIIcore}

In the strong Type II limit, $H=H(\rho)$ in the core, and invoking the
assumption that magnetic distortions are small, we can substitute the
unperturbed density of the background star, $\rho(r)$, so $H=H(r)$.

With this simplification, \eq{Ampere} is a complicated partial
differential equation even if we specify $\Jcal(A)$ somehow,
and we shall not attempt to solve it directly. Instead,
we extend the procedure first employed by
\citet{ROBERTS01081981}
for uniform density stars to stars with $H(r)\neq{\rm constant}$.
Since $B_T=0$, \eq{axiB} implies that $\Bhat\dotprod\grad A=0$ i.e.
$A$ is constant along poloidal field lines. This allows us to consider
\eq{Ampere} along individual field lines.

Let us concentrate on a particular field line. Introduce the parametric
independent variable $s$, the arc length along this field line; position
along the field line is $\vec{r}(s)$ and the tangent to the field line
is $\Bhat(s)=d\vec{r}(s)/ds$. There are two directions
normal to the field line: $\phihat$ and a second direction $\nhat(s)$
which we define to be $\nhat(s)=\phihat\crossprod\Bhat(s)$. The
field line curvature is $K(s)$ given by $K(s)\nhat(s)=d^2\vec{r}(s)/ds^2
=\Bhat(s)\dotprod\grad\Bhat(s)$. \eq{Ampere} is then equivalent to
\be
{4\pi J\over c}=4\pi \rho r\sin\theta\Jcal(A)=-\grad H\dotprod\nhat
+KH~.
\label{Amperes}
\ee
For uniform $H$ the first term in \eq{Amperes} is absent and we recover the equation
found by \citet{ROBERTS01081981}.
In \eq{Amperes} $\Jcal(A)$ is simply a constant associated with the field line.

In axisymmetry, we may choose to specify position along a field line by 
$r(s)$ and $\theta(s)$; in that case
\be
\Bhat(s)=\rhat{dr(s)\over ds}+\thetahat r(s){d\theta(s)\over ds}
\equiv\rhat\cos\Lambda(s)+\thetahat\sin\Lambda(s)
\label{Bhatspher}
\ee
and therefore
\be
\nhat(s)=\phihat\crossprod\Bhat(s)=-\rhat\sin\Lambda(s)+\thetahat\cos\Lambda(s)~.
\label{nhatspher}
\ee
The field line curvature is
\be
K(s)={d\Lambda\over ds}+{\sin\Lambda\over r}~;
\label{Kofs}
\ee
using \eq{nhatspher} and \eq{Kofs} in \eq{Amperes} we find
\be
\frac{d\Lambda}{ds}={4\pi\rho r\sin\theta\Jcal(A)\over H(r)}-{\sin\Lambda\over r}
\left(\frac{r}{H}\frac{dH(r)}{dr}+1\right)~.
\label{Lambdaeqn}
\ee
\eq{Lambdaeqn} and \eq{Bhatspher} determine the field line shape jointly, given
$\rho(r)$, $H(\rho)$ and a value of $\Jcal(A)$.

For our solutions we shall assume an $N=1$ polytrope for which $\rho(r)=\rho(0)
\sin x/x$, with $x=\pi r/R$; we shall also assume $H\propto\rho^b$. With these
choices, and the definition $ds=Rd\sigma/\pi$, \eq{Lambdaeqn} becomes
\be
\frac{d\Lambda}{d\sigma}=\frac{\Jch(A)x^b\sin\theta}{(\sin x)^{b}}
-\frac{\sin\Lambda}{x}\left[b(x\cot x-1)+1\right]
\label{dLdsig}
\ee
where 
\be
\Jch(A)={4R^2\rho(0)\Jcal(A)\over\pi H(0)}~;
\label{Jchdef}
\ee
the field line shape is found by solving \eq{dLdsig} along with
\be
\frac{dx}{d\sigma}=\cos\Lambda~~~~~x\frac{d\theta}{d\sigma}=\sin\Lambda~.
\label{xthetaeqns}
\ee
We are interested in solutions that start at some equatorial footpoint where
$x=x(0)$ and $\theta(0)=\pi/2$. At this footpoint, symmetry dictates that the
field line be perpendicular to the equator; choosing the direction to be 
vertically upward implies $\Lambda(0)=-\pi/2$.

For the ``most dipolar'' case we also want field lines to hit the outer
boundary of the Type II region at $x=x_b$ with $\Lambda=-\theta$. Very generally,
it is impossible to find solutions that are exactly vertical throughout the core
for a given $H(\rho)$. (The solutions are nevertheless close to the toy model
developed in \S\ref{toy} that demands that field lines {\sl are} 
exactly vertical.) The condition that a field line starts out at $x(0)$ and
$\theta=\pi/2$ pointing vertically and also hits vertically at $x_b$
determines the value of $\Jch(A)$ for that field line. Note that in this 
context, ``$A$'' is just a label for the field line; we could just as well
label the field line by $x(0)$. In fact, this procedure {\sl does not}
determine $A(x,\theta)$ within the Type II core. The end result is
$\Jch(x(0))$ and, since field lines hit $x_b$ at $\theta(x(0))$,
$\Jch_b(\theta)$ is determined parametrically on the outer boundary of the core. 

We can also consider other types of solution for field lines, for instance 
solutions in which field lines hit $x_b$ with some other specified set of
orientations and, therefore, carry different values of $\Jch(x(0))$ along
with them. We must be careful to choose orientations for which field lines
do not cross within the Type II core. But even that restriction allows many
different possibilities other than our ``most dipolar'' case. 

\begin{figure*}
\includegraphics[width=124mm]{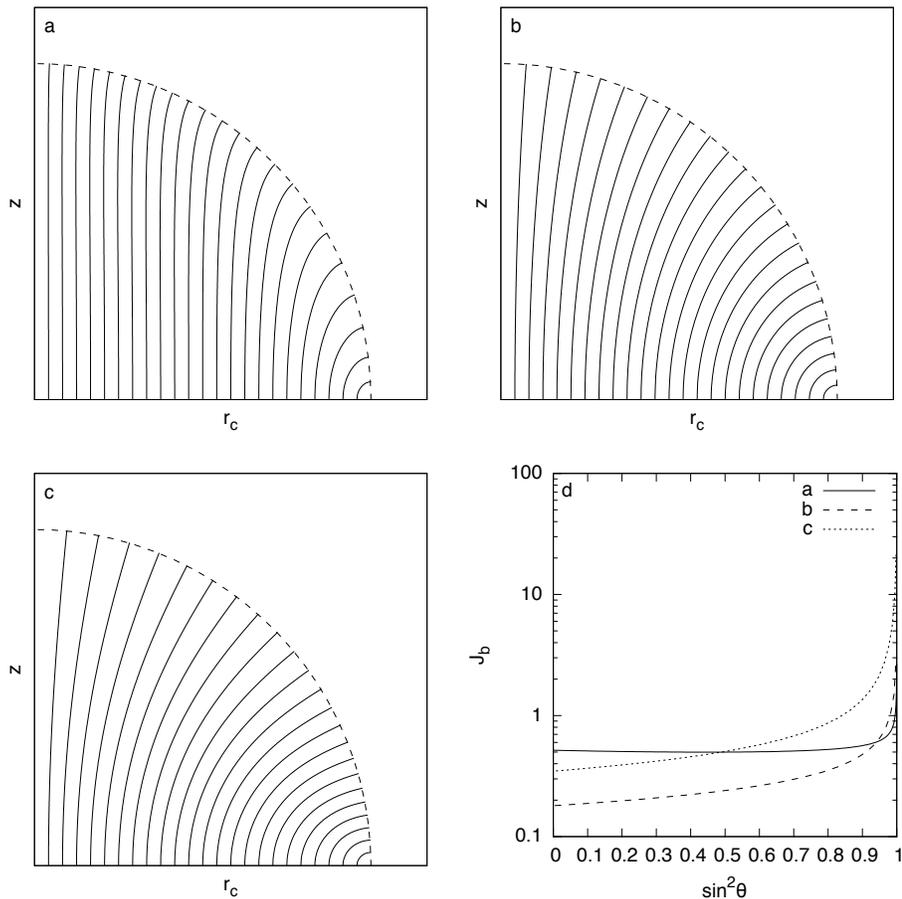}
\caption[]{{
Field line configurations in the Type II core for field lines that hit
the core boundary radially. Panel a. shows field lines for $H\propto
\rho^{1.6}$ and a $N=1$ polytrope density profile; Panel b. shows
field lines for uniform $H$ and $\rho$; Panel c. shows field lines for
uniform $H$ but a $N=1$ polytrope density profile. All three models
assume that the core radius is $r_b=0.9R$, shown on the plots as a
dashed line. Panel d. shows the rescaled current parameter $\Jhat$ for
the three models as a function of $\sin^2\theta$ on $r_b$.}
\label{fig:Field-radial}}
\end{figure*}

{As an illustration, Fig. \ref{fig:Field-radial} shows the field lines assuming
that field lines intersect the outer boundary of the core {\sl radially}. For
$H_b\gg B$, this is the field configuration needed to match smoothly into
regions without currents since $\vert B_\theta\vert\sim\vert B_r\vert$ in
that case \citep[e.g.][]{ROBERTS01081981}. The current free region may be 
the normal shell or the vacuum exterior. In order to highlight the effects
of variable $H$ and $\rho$ inside the core, we show three different cases:
(a) $H\propto
\rho^{1.6}$ with a $N=1$ polytrope density profile;
(b) $H$ and $\rho$ uniform, the case considered by \citet{ROBERTS01081981};
(c) $H$ uniform but with a $N=1$ polytrope density profile. The lower right
panel of Fig. \ref{fig:Field-radial}
shows $\Jch_b$ on the boundary of the Type II
core, which we placed at $r_b=0.9R$, as a function of $\sin^2\theta$. 
As can be seen from 
Fig. \ref{fig:Field-radial}
the current diverges for field lines near the equator. This is because they
start out vertical and must turn through about $\pi/2$
to hit the boundary radially within the short distance $r_b-r$.}

The fact that we do not determine $A(x,\theta)$ in the Type II core means
that even with a given set of solutions for individual field lines, we can
consider various ``field line densities'' within the core. This freedom
can accommodate theories of field line evolution in which secular motion
of flux tubes leads to different concentrations of flux within the core.

\subsection{Solution in the Normal Shell}
\label{normalshell}

Within the normal shell, it is most straightforward to solve \eq{Ampere}
directly: using $\Hvec=\Bvec$ and substituting for the poloidal field we get
\be
-{H(0)R^2\over\pi^2}[\Jch(A)x\sin x\sin^2\theta]
={\partial^2A\over\partial x^2}+{1\over x^2}\left({\partial^2A\over\partial
\theta^2}-\cot\theta{\partial A\over\partial\theta}\right)~,
\label{normalshelleqn}
\ee
where we have used the same parametrization as in \eq{Jchdef} and specialized
to the $N=1$ polytrope density profile. \eq{normalshelleqn} would be straightforward
to solve {\sl given} $\Jch(A)$. However, we have to determine $\Jch(A)$
to be consistent with matching conditions at both the inner boundary of the
normal shell at $x_b$ and at the stellar surface, where the field must match
smoothly to vacuum. 

Matching to a vacuum exterior is accomplished most easily by introducing
the expansion
\be
A(x,\theta)=\sin\theta\sum_{j=0}^\infty A_j(x)\pj(\cos\theta)
\label{Aexpand}
\ee
where $\pj(\mu)$ is an associated Legendre polynomial. Using \eq{Aexpand}
in \eq{normalshelleqn} implies
\be
\frac{d^2A_j(x)}{dx^2}-\frac{(2j+2)(2j+1)A_j(x)}{x^2}
=-\frac{H(0)R^2x\sin x}{\pi^2N_j}\int_{-1}^{+1}d\mu\sqrt{1-\mu^2}\pj(\mu)
\Jch(A)
\label{Ajeqn}
\ee
where 
\be
N_j={2(2j+2)(2j+1)\over 4j+3}~.
\label{Njeqn}
\ee
The source term in \eq{Ajeqn} generally involves the entire set of $A_i$,
not just $i=j$. But once $\Jch(A)$ has been determined, the source term
is known, and the boundary condition at the surface $x=\pi$ is
\be
\left(\frac{dA_j(x)}{dx}\right)_\pi+\frac{(2j+1)A_j(\pi)}{\pi}=0
\label{surfacebc}
\ee
for matching to a vacuum field in the exterior.

Matching to the core requires that the normal component of $\vec{B}$ and
tangential component of $\vec{H}$ be continuous across the boundary. In 
the simplest case, the transition zone thickness may be neglected. 
If $\Lambda_b(\theta)$ is the value of $\Lambda$ for a field line hitting
$x_b$ at $\theta$ and $B(\theta)$ is the magnitude of the core induction
field there, then the matching conditions are ($\mu=\cos\theta$)
\ba
\left(\frac{\partial A}{\partial\mu}\right)_b&=&-\frac{R^2x_b^2B(\theta)
\cos\Lambda_b(\theta)}{\pi^2}
\nonumber\\
\left(\frac{\partial A}{\partial x}\right)_b&=&-\frac{R^2x_bH_b\sin\theta
\sin\Lambda_b(\theta)}{\pi^2}~.
\label{bcszeroell}
\ea
In the strong Type II regime, $B(\theta)\ll H_b$ and unless 
$\Lambda_b(\theta)\lesssim B/H_b$ the field is predominantly tangential
at the base of the normal shell. To a first approximation, $(\partial A/
\partial\mu)_b\approx 0$, which is equivalent to $A_j(x_b)\approx 0$,
in the strong Type II regime.

\eq{surfacebc} and \eq{bcszeroell} impose three conditions on each
$A_j(x)$, which is one too many. This overdetermination implies constraints
on the form of $\Jch(A)$ within the shell.  

In the ``most dipolar'' case, $\Lambda_b(\theta)=-\theta$, and
\eq{bcszeroell} implies that the field at the inner edge of the normal
shell is approximately $-\thetahat H_b\sin\theta$ in the limit $B\to 0$.
In this limit, field lines from the interior do not penetrate into the
normal shell but still influence conditions there via \eq{bcszeroell}.
With $B=0$, the field inside the normal shell is precisely dipolar,
$A(x,\theta)=A_0(x)\sin^2\theta$, where
\ba
A_0(x)&=&\frac{H_bR^2f(x;x_b)}{3\pi^2}
\nonumber\\
f(x;x_b)&=&\frac{x^3-x_b^3+6x\cos x-6x_b\cos x_b+(3x^2-6)\sin
x-(3x_b^2-6)\sin x_b}{x(1+\cos x_b)}~.
\label{mostdipolar}
\ea
Within the shell, consistency with \eq{surfacebc} and \eq{bcszeroell} requires that
\be
\Jch(A)=\Jchshell=\frac{H_b}{H(0)(1+\cos x_b)}=\frac{(\sin x_b/x_b)^{b}}{1+\cos x_b}
\label{jchshell}
\ee
which is the simplest case showing how the boundary conditions determine $\Jch(A)$.
The stellar dipole moment is 
\be
\mu=RA_0(\pi)={H_bR^3f(\pi;x_b)\over 3\pi^2}~;
\label{dipolemoment}
\ee
if $\pi-x_b\equiv\delb=\pi\epsb\ll 1$,
$f(\pi; x_b)/3\pi^2\approx\delb/3\pi=\epsb/3$, which leads to 
\eq{dipolefieldstrength} for thin shells.

\eq{mostdipolar} holds when we can neglect the thickness of the transition layer and
the radial field that pokes into the normal shell through it. If the transition layer
is thin enough, then its effect is to introduce a ``surface current'' into the jump
condition from the Type II core to the normal shell: the tangential field at the base
of the normal shell is now
\be
B_\theta(x_b,\theta)=-\frac{\pi^2}{R^2x_b\sin\theta}\left(\frac{\partial A}{\partial x}
\right)_b=H_b\sin\Lambda_b(\theta)+\frac{\pi\ell\Jch_b(\theta)H(0)\sin x_b\sin\theta}{R}
\label{jump}
\ee
where $\ell$ is the thickness of the transition zone. \S\ref{transition} contains a brief
discussion of the rather complex physics of this region. There, it is shown that \eq{jump}
holds as long as field lines do not rotate very much as they pass through the transition
zone. Roughly speaking, this will be true provided that $H_b\ell/BR\ll 1$; therefore
the second term in \eq{jump} is $\ll BH(0)/H_b^2$ times the first.
Nevertheless, \eq{jump} perturbs the solution away from \eq{mostdipolar} even
in the ``most dipolar'' case. In particular, the jump term engenders non-dipolar fields
in the shell and outside, and requires a more detailed calculation of $\Jch(A)$ inside
the shell. As long as the corrections to \eq{mostdipolar} are small they may be
computed perturbatively. A method of calculation is outlined in Appendix 
\ref{aperturbsoln}.

\begin{figure*}
\includegraphics[width=124mm]{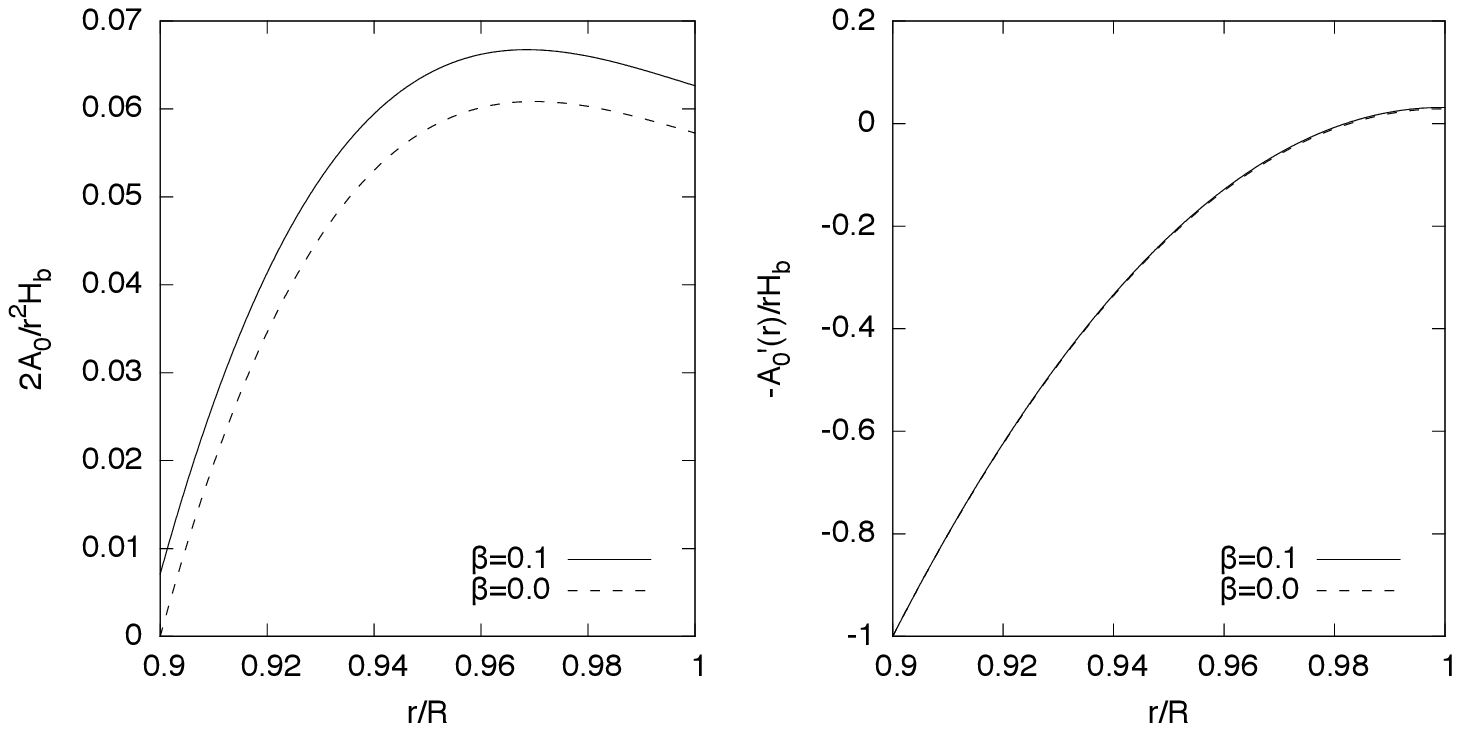}
\caption[]{{$2A_0/H_br^2$ (left panel) and $-(H_br)^{-1}dA_0/dr$ (right panel) in
the normal shell, where $\Hvec=\Bvec=\rhat(2A_0/r^2)\cos\theta-(r^{-1}dA_0/dr)\thetahat
\sin\theta$. For the functional form of $A_0$, see \eq{AzB}.
Two different values of $B$ are assumed, parameterized by
the quantity $\beta$ defined in \eq{betaftildefs}; results are shown for 
$\beta=0$ and $0.1$.}
\label{fig:Field-inshell}}
\end{figure*}

Relaxing the $B\to 0$ limit also perturbs the solution even if the transition zone
thickness is negligible. If we assume that 
\be
A(x_b,\theta)=\onehalf Br_b^2\sin^2\theta=\frac{BR^2x_b^2\sin^2\theta}{2\pi^2}
\ee
then the unperturbed field inside the normal shell remains dipolar, and is altered
to $A_0(x,\theta)=A_0(x;B)\sin^2\theta$, where
\be
A_{0}(x;B)=\frac{BR^2x_b^3}{2\pi^2 x}+\frac{(H_b+B/2)R^2f(x;x_b)}{3\pi^2}
\label{AzB}
\ee
and the current density parameter is
\be
\Jchshell=\frac{H_b(1+B/2H_b)}{H(0)(1+\cos x_b)}~.
\label{Jchshellgen}
\ee
\eq{AzB} implies a magnetic moment
\be
\mu(B)={Br_b^3\over 2}+{(H_b+B/2)R^3f(\pi;x_b)\over 3\pi^2}
\label{dipolemomB}
\ee
which reduces to \eq{dipolemoment} for $B=0$;
the first term in \eq{dipolemomB} is simply the dipole moment associated 
with $B$ and the second arises from the currents in the normal shell. 
{Fig. \ref{fig:Field-inshell} shows $A_0/r^2$ and $-r^{-1}dA_0/dr$
in the normal shell, where the magnetic field is
 $\Hvec=\Bvec=\rhat(2A_0/r^2)\cos\theta-(r^{-1}dA_0/dr)\thetahat
\sin\theta$. In the limit that the transition from superconductor
to normal conductor takes place within a region of zero thickness,
$H$ is discontinuous at the boundary between core and shell, but 
the radial component of $\Bvec$ and the $\thetahat$
component of $\Hvec$ are continuous, as required by Maxwell's equations
with no surface currents. Appendix \ref{transition} outlines how $\Hvec$ field changes 
smoothly when the transition region has nonzero thickness. 
Within the Type II core, $H=H(\rho)$ is independent
of $\theta$.}

\eq{AzB} implies that magnetic field lines from the core enter the shell. The
bounding field line entering from the core has $A=\onehalf Br_b^2$, and therefore
defines a region $\theta\leq\thm$ where
\be
\sin^2\thm={\beta\over\beta x_b/x+\ftil(x;x_b)}~;
\label{boundingfieldline}
\ee
in \eq{boundingfieldline}
\be
\beta\equiv\frac{3\pi^2Br_b^2}{(2H_b+B)R^2f(\pi;x_b)}~~~~~~
\ftil(x;x_b)\equiv\frac{f(x;x_b)}{f(\pi;x_b)}~.
\label{betaftildefs}
\ee
The region occupied by the field lines is small if $\beta\sim B/H_b\epsb\ll 1$;
note that this is more restrictive than the condition for validity of the strong Type II
regime, $B\ll H_b$. 
\eq{boundingfieldline} indicates that the bounding field line rises from 
$\theta=\pi/2$ close to $x_b$, remaining at $\theta\sim 1$ as long as $\ftil(x;x_b)
\sim\beta$, which is the case for $x/x_b-1\lesssim\beta$. Ultimately, the bounding
field lines emerge at $x=\pi$ confined to a polar cap with $\theta\lesssim\sqrt{\beta}$.
The volume of the region occupied by impinging field lines is only $\sim\beta$ times
the volume of the shell. In the limit $\beta\to 0$ field lines from the core 
{\sl do not} penetrate into the normal shell.

However, within that volume, $\dJch(A)=\Jch(A)-\Jchshell\neq 0$ because field 
lines entering from the core carry their $\Jch(A)$ along with them. The zeroth order
solution represented by \eq{AzB} and \eq{Jchshellgen} requires perturbative corrections
driven by the difference $\dJch(A)$ within this region; the 
corrections are ${\cal O}(\beta\dJch(A)/\Jchshell)$.
Appendix \ref{aperturbsoln} outlines how these corrections may
be taken into account. 

\subsection{Magnetic Distortions Due to Poloidal Fields}
\label{magdistort}

For poloidal fields, the Lorentz acceleration is $-\grad\Phi=\Jcal(A)\grad A$,
so
\be
\Phi(A)=-{\pi H(0)\over 4R^2\rho(0)}\int_0^A dA'\Jch(A')~,
\label{phimdef}
\ee
and the Bernoulli equation is 
\be
C=h+\Psi+\frac{\partial f}{\partial\rho}+\Phi~
\label{Bernoulli}
\ee
where $C$ is the Bernoulli constant. For perturbations of a $N=1$ polytrope, if
we expand in Legendre polynomials 
\be
\Psi(x,\theta)=\sum_l \Psi_l(x)P_l(\cos\theta)
\label{psiexpand}
\ee
we get
\be
\frac{1}{x^2}\frac{d}{dx}\left[x^2\frac{d\Psi_l(x)}{dx}\right]+
\left[1-\frac{l(l+1)}{x^2}\right]\Psi_l(x)=\delta C\delta_{l,0}-
\left(\frac{\partial f}{\partial\rho}+\Phi\right)_l\equiv \delta C\delta_{l,0}+M_l(x)
\label{psileqn}
\ee
where the magnetic potentials are expanded in the same fashion as 
\eq{psiexpand} and $\delta C$ is the perturbation of the Bernoulli
constant away from its background value. The general solution of 
\eq{psileqn} that is regular at $x=0$ is
\be
\Psi_l(x)=\delta C\delta_{l,0}+j_l(x)\left[K_l+\int_x^\pi dx'(x')^2y_l(x')M_l(x')
\right]+y_l(x)\int_0^x dx'(x')^2j_l(x')M_l(x')
\label{psilsoln}
\ee
where $j_l(x)$ and $y_l(x)$ are spherical Bessel functions. The constants
$K_0$ and $\delta C$ are determined by the conditions $\Psi_0(\pi)=0=
[d\Psi(x)/dx]_\pi$, which imply
\be
\delta C=K_0=-{1\over\pi}\int_0^\pi dx x^2 j_0(x)M_0(x)~,
\ee
and the constants $K_l$ are determined by the conditions $[d\Psi_l(x)/dx]_\pi
+(l+1)\Psi_l(\pi)/\pi=0$, which imply ($f_l'(x)\equiv df_l(x)/dx$)
\be
0=K_l\left[j_l'(\pi)+\frac{(l+1)j_l(\pi)}{\pi}\right]
+\left[y_l'(\pi)+\frac{(l+1)y_l(\pi)}{\pi}\right]
\int_0^\pi dx x^2j_l(x)M_l(x)~.
\label{Kleqn}
\ee
For $l>0$, 
\be
\Psi_l(\pi)=-\frac{1}{\pi^2j_l'(\pi)+(l+1)\pi j_l(\pi)}
\int_0^\pi dx x^2j_l(x)M_l(x)~.
\label{Psilpi}
\ee
We may define the stellar multipole moment by $\Psi_l(\pi)=-GM\Mcal_l/R$.
With this definition, the contribution to the moment of inertia tensor
resulting from the quadrupolar distortion $\Mcal_2$ is
$\delta{\mathbf I}_2=-\onethird MR^2\Mcal_2(-\xhat\xhat-\yhat\yhat+2\zhat\zhat)$.

In the ``most dipolar'' case with negligible transition layer thickness
and core magnetic induction, the distortion arises entirely from magnetic
stresses inside the crust and is quadrupolar. In this case, ($\Theta(z)=1$
for $z>0$, zero otherwise)
\be
M_2(x)=-\frac{H_b^2f(x;x_b)\Theta(x-x_b)}{18\pi\rho(0)(1+\cos x_b)}
=-\frac{2H_b^2R^3f(x;x_b)\Theta(x-x_b)}
{9\pi^2M(1+\cos x_b)}
\ee
and ($\pi^2j_2'(\pi)+3\pi j_2(\pi)=\pi$)
\be
\Mcal_2=-\frac{2H_b^2R^4}{9\pi^3GM^2}\int_{x_b}^\pi\frac{dx x^2 j_2(x) f(x;x_b)}
{1+\cos x_b}\equiv-\frac{2H_b^2R^4\Ical(x_b)}{9\pi^3GM^2}~.
\label{quadrupolemoment}
\ee
For thin shells, $\Ical(x_b)\approx 9\pi/2=14.14$, independent 
of the shell thickness; for $\epsb=0.1$, it is $\Ical(0.9\pi)=11.44$;
consequently
\be
\Mcal_2=-1.39\times 10^{-9}H_{b,14}^2R_{10}^4M_{1.4}^{-2}\left[\frac
{\Ical(x_b)}{10}\right]
\label{quadrupolemomentevaluated}
\ee
where $H_b=10^{14}H_{b,14}$ G, $R=10R_{10}$ km and $M=1.4M_{1.4}\msun$.
{Thus, the scale of the quadrupolar distortion is determined by the value
of $H_b$ for $B/H_b\ll 1$; for terms $\mathcal{O}(B)$, see \eq{quadmomexpand}
and \eq{Ccoeffs}.}


\def\Jchz{\Jch^{(0)}}
\def\qcore{q_{2,{\rm II}}}
\def\qshell{q_{2,{\rm shell}}}
\def\Qshell{Q_{2,{\rm shell}}}

\section{Toy Model: Vertical Field in the Core}
\label{toy}

We shall be interested in solutions that lead to external fields as
close to dipolar as possible; such solutions have fields that are
vertical at the outer boundary of the Type II core. In general, these
will not allow field lines that are exactly vertical throughout the
core.

Nevertheless, to develop a toy model,
suppose $\Hvec=H(r)\zhat$; this model will illustrate many features
of the problem. For vertical fields, $\Jcal_{\rm II}(A)={\rm constant}=
\Jcal_{\rm II}$ and from Amp\'ere's law
\be
H(r)={4R^2\rho(0)\Jcal_{\rm II}(1+\cos x)\over\pi}=\onehalf H(0)(1+\cos x)
\label{tosoln}
\ee
where $x=\pi r/R$ and $H(R)=0$. The dimensionless combination
\be
\Jchc\equiv {4R^2\rho(0)\Jcal_{\rm II}\over\pi H(0)}=\onehalf~;
\label{Jcoretoy}
\ee
moreover
\be
b={d\ln H\over d\ln\rho}={x\sin x\over (1+\cos x)(1-x\cot x)}
\label{dlnHtoy}
\ee
varies monotonically between $b=3/2$ and $b=2$ from the center of
the star to the surface. Although this model is not truly physical
because $H(\rho)=H(\rho/\rho(0))$ depends on the central density,
hence mass, of the star and so involves a density scale that is 
not the same for all stars, $3/2\leq b\leq 2$ is an acceptable
range for nuclear equations of state. Moreover, when we consider
models with $H\propto\rho^b$ with realistic values of $b$, we shall see
that even though $\Jch(A)$ is no longer constant, its values are
close to $0.5$ for cases where field lines hit the outer boundary
of the core vertically; because field lines are vertical at the
equator, they remain close to vertical throughout the core in these
cases, as may be seen from Fig. \ref{fig:Field-config}.

%
The interior field must match onto the field in the normal shell.
If the transition from superconductor to normal conductor occurs
in a transition region of negligible thickness, then
%
the field is dipolar in the shell to a good approximation,
and is given by \eq{AzB} and \eq{Jchshellgen}.
\eq{Jchshellgen} holds generally for the dipole field within 
the normal shell; for the toy model $H_b=\onehalf H(0)(1+\cos x_b)$ so
\be
\Jchshell=\onehalf\left(1+{B\over 2H_b}
\right)=\Jchc\left(1+{B\over 2H_b}\right)~.
\label{Jshelltoy}
\ee
The core and shell current density parameters are nearly the same in this
toy model for $B/2H_b\ll 1$. This will also be true for more realistic
models, but the difference will be $\sim 0.1$, not $\sim B/H_b$. 
For the toy model, the perturbations caused by $\dJch(A)$ are 
very small, since $\beta(\Jchc-\Jchshell)=\onehalf\beta B\Jchc/H_b$,
and we shall ignore them; for realistic models, where $\Jchc-\Jchshell
\sim 0.1$, the perturbations are larger, $\sim 0.1\beta$.

In general, there will be a contribution to the stellar distortion 
$\propto B$. One reason is that $H_b\to H_b+B/2$
in the dipole solution within the shell; thus we substitute 
$H_b^2\to (H_b+B/2)^2\simeq H_b^2+H_bB$ in \eq{quadrupolemoment}.
There are also contributions from the terms in the solution that
are $\propto B$.
For the toy model with $\Jchc=\onehalf\simeq\Jchshell$
the extra $l=2$ contribution to the driving term is
\be
\Delta M_2(x)=-\frac{\pi H(0)Br_b^2}{24R^2\rho(0)}\left\{
\begin{array}{ll}
(x/x_b)^2   &\mbox{if $x\leq x_b$}\\
x_b/x&\mbox{if $x_b\leq x\leq\pi$}
\end{array}
\right.
\ee
which implies an additional quadrupole moment 
\be 
\Delta\Mcal_2=-\frac{H(0)Br_b^2R^2\dIcal(x_b)}{6\pi GM^2}
=-\frac{1.03\times 10^{-9}H_{b,14}^2R_{10}^4\dIcal(x_b)}
{M_{1.4}^2}
\,\frac{B}{H_b}
\label{dquadrupolemoment}
\ee
where $\dIcal(x_b)=\dIcalc(x_b)+\dIcals(x_b)$ with core and shell contributions
\ba
\dIcalc(x_b)&=&\frac{2(r_b/R)^2}{1+\cos x_b}
\left[\left(\frac{15}{x_b^2}-6\right)\sin x_b+
\left(x_b-\frac{15}{x_b}\right)\cos x_b\right]
\nonumber\\
\dIcals(x_b)&=&\frac{2(r_b/R)^2}{1+\cos x_b}
\left[-x_b(1+\cos x_b)+3\sin x_b\right]
\label{dIcals}
\ea
respectively.
The crust contribution $\dIcals(x_b)$
is the same in realistic models.
For thin shells, $\dIcals(x_b)
\simeq 12/\delb$ 
and $\dIcals(0.9\pi)=26.1$. The core contribution
$\dIcalc(x_b)$ will be close to the value in the toy model, with differences
arising because field lines are not strictly vertical, realistically. For
thin shells, $\dIcalc(x_b)\simeq (60/\pi-4\pi)/\delb^2\simeq 6.5/\delb^2$ 
in the toy model and $\dIcalc(0.9\pi)=35.8$.
Comparing \eq{dquadrupolemoment} with \eq{quadrupolemomentevaluated}
we see that $\vert\Delta\Mcal_2\vert\lesssim(\beta/\epsb)\vert\Mcal_2\vert$ 
characteristically. 



\section{Numerical Solutions}
\label{solutions}

\subsection{Field Line Structure}

In the strong Type II limit where $H=\Hc(\rho)$ does not depend on $B$,
we can solve \eq{dLdsig} and \eq{xthetaeqns} for each magnetic field line
individually. Symmetry dictates that field lines be vertical at their
equatorial footpoints. For the ``most dipolar'' solutions that match
to a pure dipole field in the normal shell to zeroth order in $\ell$
and $\beta$, we also require that each field line be vertical at $x_b$.
We can label each field line by its footpoint radius, $x_0$. Once we
specify $H(\rho)=\Hc(\rho)$ we can find the value of $\Jch(x_0)$ for 
which these conditions are satisfied. Thus, we determine both the source
and the field line shapes that are consistent with our boundary conditions.

In order to choose the function $H(\rho)$ we fit the values of the proton
fraction given in \citet{PhysRevC.78.015805} to power laws of the baryon
density. Writing $H(\rho)\propto\rho^b$, we found that 
$b\approx1.6-1.8$. We used the $N=1$ polytrope for the mass density
profile of the star.

\begin{figure*}
\includegraphics[width=124mm]{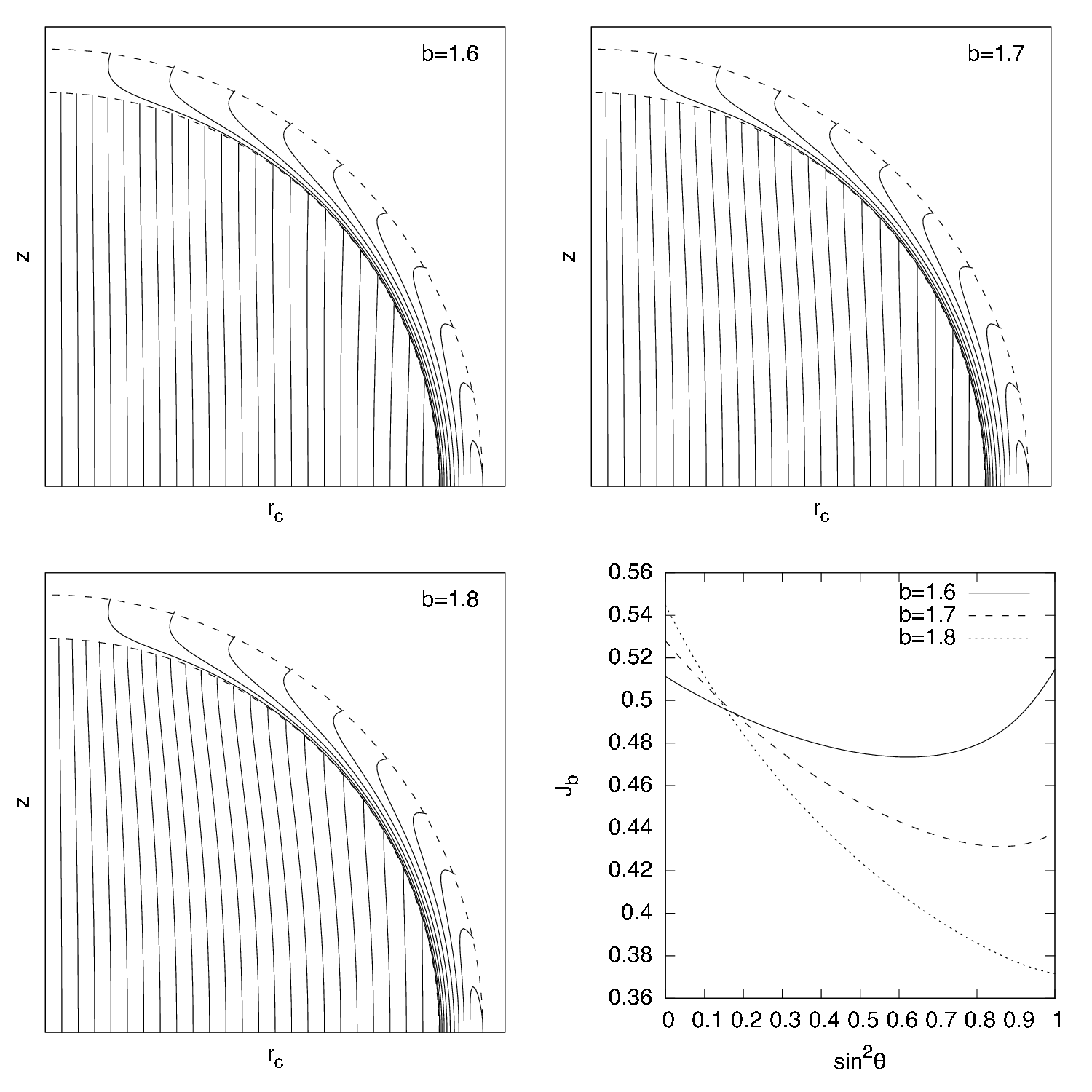}
\caption[]{{Field line configurations assuming $H(\rho)\propto\rho^b$ with $b=1.6,\,1.7$ and
$1.8$. The inner and outer dashed lines show the boundary of the
Type II core at $r=0.9R$ and the stellar surface at $r=R$, respectively, for the
stellar magnetic axis in the $z$ direction. The lower right panel shows
the value of the rescaled current parameter $\Jhat$ as a function of
$\sin^2\theta$ along the inner dashed boundary.}
\label{fig:Field-config}}
\end{figure*}

Fig. \ref{fig:Field-config} shows the results for $b=1.6,\,1.7$ and $1.8$.
Note that there are two competing features of the solutions: $\grad H$ tends
to curve the field lines inward, and $\Jch(x_0)$ opposes this trend. The
field lines are most nearly vertical throughout the core for $b=1.6$, the
smallest value displayed. Even though the field line shapes are not
precisely vertical throughout the core, the deviations are relatively small.
We have also found the field line solutions for $b=1,\,1.5,$ and $2$; they
are substantially similar to the configurations shown in Fig. \ref{fig:Field-config}.

The lower right panel shows the current density parameter $\Jch_b$ as a function
of $\sin^2\theta$. This quantity is found by determining $\Jch(x_0)$ for each
field line; using the solution for $\theta(\sigma)$ along the line, we find 
where it intersects the surface $x_b$. Note that our solutions {\sl do not}
determine the value of $A$ for a given field line.

For $B=0=\beta$ and $\ell=0$ boundary conditions at the core-shell interface
dictate that $\Bvec=-\thetahat H_b\sin\theta$ at the base of the shell. This
means that in this approximation field lines {\sl do not} penetrate from the
core to the shell. The solution inside the shell that matches smoothly to 
a vacuum field is given by \eq{mostdipolar}. Since there are two boundary
conditions at $x_b$, the surface boundary condition determines $\Jchshell$
(\eq{jchshell})
so the entire solution outside the core not only is determined fully but 
also depends on $H_b$.

\begin{figure*}
\includegraphics[width=124mm]{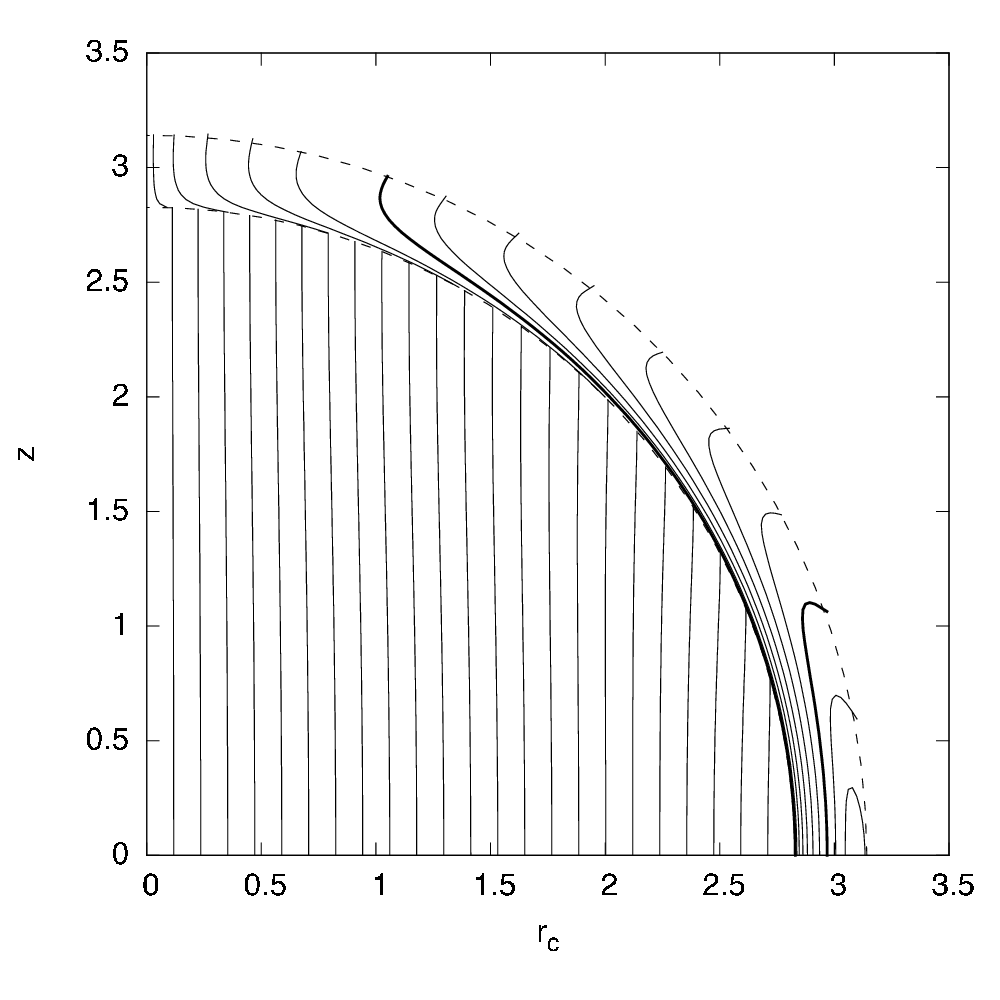}
\caption{{
Field line configuration in the same layout as in Fig. \ref{fig:Field-config}, with nonzero $B$
field strength at the core boundary. Shown here is the case where
$\beta=0.1$, as defined in \eq{betaftildefs}, and $b=1.6$. Field lines in the
shell are computed to lowest order in $\beta$; for corrections see \S \ref{perturbations}.}
\label{fig:Field-finiteB}}
\end{figure*}

Fig. \ref{fig:Field-finiteB} illustrates the field line configuration for $b=1.6$ but
$\beta=0.1$, to lowest order in $\beta$. (Corrections will be considered in \S\ref{perturbations}.)
In this case, field lines from the core penetrate into the normal shell. The inner
thick line shows the bounding field line. It is apparent that field lines hug close to
$x=x_b$ before fanning out into a polar cap whose angular extent is $\sim\sqrt{\beta}$.
Also shown in this figure is the small region where field lines in the shell are closed: the
first open field line has its inner footpoint at $x\simeq 0.97\pi$.
(This region is also apparent in the $\beta=0$ configurations in Fig. \ref{fig:Field-config}.)
In a fluid shell, this region would be magnetohydrodynamically unstable
\citep{1973MNRAS.162..339W,1973MNRAS.163...77M} but shear stresses in the normal shell restore stability since the
shear modulus $\mu\gg B^2\sim 10^{24}B_{12}^2\,{\rm g\,cm^{-1}s^{-2}}$
\citep[e.g.][]{2001LNP...578..127H, 2009PhRvL.102s1102H, 2011MNRAS.416...22B}.

\subsection{Magnetic Distortion: Mass Quadrupole Moment}
\label{quadrupoledistortion}

In general, there are four contributions to the quadrupolar distortion
to linear order in $B/H_b$:  
\be
\Mcal_2=-10^{-9}H_{b,14}^2R_{10}^2M_{1.4}^{-2}
\left[C_1+(C_2+C_3+C_4)\frac{B}{H_b}\right]~.
\label{quadmomexpand}
\ee
The coefficients $C_1$ and $C_2$ 
arise from the shell, and are independent of the core structure; for
$r_b=0.9R$, $(C_1,C_2)=(+1.58,+28.3)$.
The coefficient
$C_3$ arises from $\Phi(A)$ and the coefficient $C_4$ arises from 
$B\partial f/\partial\rho$; these depend on the structure of the core
field: for $r_b=0.9R$
\be
(C_3,C_4)=\left\{\begin{array}{ll}
(+30.4,-0.434)   &\mbox{if $b=1.6$}\\
(+35.4,-2.20)   &\mbox{if $b=1.7$}\\
(+40.7,-4.64)   &\mbox{if $b=1.8$}\\
(+36.7,0)        &\mbox{Toy}
\end{array}
\right.
\label{Ccoeffs}
\ee
where in all cases $C_3>\vert C_4\vert$ because field lines are
nearly vertical (exactly vertical in the toy model). Roughly speaking,
the coefficient of $B/H_b$ in \eq{quadmomexpand} is $\simeq 60$,
but $B/H_b\approx2f(\pi;x_b)R^2\beta/3\pi^2r_b^2=\beta/14.1$ for $r_b=0.9R$,
which mitigates the dependence of $\Mcal_2$ on $B$ for small $\beta$.


\subsection{Perturbative Corrections: Nonzero $\ell$ and $\beta$}
\label{perturbations}

Table \ref{tab:shell-values} shows the results of computing perturbations
resulting from nonzero transition zone thickness using the method
developed in Appendix \ref{transitionfield}. Although the table only
contains results for multipoles $Q_j$ with $j\leq 5$, the calculations
were performed using $\Jhat_{b,j}$ with $j\leq 14$, and therefore
determined $Q_j$ for $j\leq 14$. The procedure can be iterated but
we did not do so. To obtain these results, we solved \eq{findJn}
with $\Delta A_j(x_b)=0$. Appendix \ref{transition} shows that in fact
$\Delta A_j(x_b)\neq 0$. However, according to \eq{changeinAlayer},
$\Delta A_j(x_b)$ is primarily dipolar to $\mathcal{O}(\ell)$. This
dipolar component as well as the effect of $\Jhat_{b,0}$ may be
absorbed into a ``renormalized'' background solution. The second
line in Table \ref{tab:shell-values} tabulates the ``extra'' change
in $Q_0$ that results when we renormalize in this fashion.
\begin{table}
\begin{tabular}{ccc}
\hline
$j$ & $\Jhat_{b,j}$ & $Q_j(\ell)/R^{2j-1}\ell\mu$
\\ \hline
 0 & -4.918e-01 & 1.701e+01  \\ 
 0 & 0 & 5.239e-01  \\ 
 1 & 6.067e-03&  -4.999e-01  \\ 
 2 & -5.198e-03 & 2.723e-01  \\ 
 3 & 1.241e-03 & -6.844e-02 \\ 
 4 &  -3.843e-04 & 1.973e-02 \\ 
 5 &   8.526e-05& -3.657e-03 \\ 
\hline
\end{tabular}
\caption{Values of $\Jhat_{b,j}$ (col. 2), $Q_j(\ell)/R^{2j}\ell\mu$ (col. 3)
for $j\leq 5$ (col. 1)
and $b=1.6$ and $r_b=0.9R$. Second line renormalizes $H_b$ as described in
the text.}
\label{tab:shell-values}
\end{table}

Table \ref{tab:shell-values} shows that the multipole perturbations that result
from nonzero $\ell$ are small but not entirely negligible. For example, the
shift in the dipole moment is $\Delta\mu/\mu=-17.01\ell/R$; for $\ell/R=0.01$
this is a downward change of about 17\%. Since we require $\Delta A_j(x_b)$ to be
relatively small, \eq{changeinAlayer} indicates that $\ell/R\ll B/H_b\sim\beta
\epsb$, so consistency would require somewhat smaller $-\Delta\mu/\mu$:
shifts are restricted to $\lesssim$ a few percent. The other multipole moments 
are smaller by at least a factor of 30.

Implementing the procedure in Appendix \ref{dipolepokesin} proved more challenging
computationally, particularly for larger values of $\beta$, where the various
matrices that had to be inverted turned out to be nearly singular. These
computational problems did not arise for $\beta\leq 0.05$, and for $10^{-5}\leq
\beta\leq 0.05$ we found $\Delta\mu/\mu\simeq-0.115\beta$ and a RMS change 
$\Delta A_{\rm rms}(R)/A_0(R)\simeq 0.150\beta$ assuming $b=1.6$ and $\epsb
=0.1$. Thus, although the perturbations 
are predominantly dipolar, nondipole contributions are only a few times smaller. 
These results were found using various maximum $0\leq j_{\rm in}\leq 10$ in the expansion 
for $\Jhat_b(\theta)$, and various maximum $10\leq j_{\rm max}\leq 14$ in the 
multipole expansion within the normal shell. 
Empirically, we found that about 90\% of the perturbation arose when we took 
$j_{\rm in}=0$, which suggests that the main effect arises from the peculiar
shape of the region within the shell occupied by field lines entering from
the Type II core; see Fig. \ref{fig:Field-finiteB}.

\section{Discussion and Conclusions}
\label{conclusions}

In this paper, we developed the formalism for computing the configuration of poloidal
magnetic fields inside a neutron star with a strongly Type II superconducting core.
Qualitatively, the restriction to hydrostatic balance without ``magnetic buoyancy''
imposes constraints on the electric current density given by \eq{jcalA}. Because of 
these constraints, we must determine both the current densities and field line 
structure at the same time. Within the Type II core, the task is simplified when 
$B\ll H$, for in this limit $H$ is a function of density, and therefore a function 
of radius, to lowest order in the magnetic distortion.

The main results of this paper are:
\begin{enumerate}
\item The Type II core provides an important boundary condition for fields in the 
normal shell surrounding it; see \eq{bcszeroell}.
\item If the core is strongly Type II, continuity of the tangential and normal components
of magnetic field imply that the field points very nearly along the inner boundary of
the normal shell, unless the field just inside the boundary is tuned finely. Thus,
generically field lines from the core do not penetrate deep into the normal shell of
the star. However, the field strength at the base of the normal shell is $\sim H_b$ in
this case.
\item Substantial currents in the normal shell are required for the field to adjust in
strength and direction to conform with vacuum boundary conditions at the radius of the
star.
\item In the ``most dipolar case'' in which the field is vertical at the outer edge of
the Type II core, the emerging field is dipolar, with a surface field strength given
by \eq{mostdipolar}.
\item If the thickness of the normal shell of the star is $\epsb R$ with $\epsb\ll 1$,
the surface field strength for the ``most dipolar'' case 
is $\simeq H_b\epsb/3$. For typical neutron star parameters
\eq{dipolefieldstrength} implies a surface field strength
$\simeq 3\times 10^{12}$ G, which is characteristic of the large population
of ``normal neutron stars.''
\item Magnetic fields induce a quadrupolar distortion of the star $\mathcal{M}_2\simeq -10^{-9}$.  
The scale of the distortion is set primarily by $H_b$ for small $B/H_b$; see \eq{quadmomexpand}.
\item Perturbations arise from the finite thickness of the transition layer in which
the superconductor disappears and from field lines that penetrate from the Type II core
to the normal shell. These perturbations are small if $\ell/\epsb R\ll\beta\ll 1$, which
delineates the domain of validity of our calculations.
\end{enumerate}
Although our calculations assume a star in hydrostatic balance, 
some of the qualitative features found here
would hold for stars that are out of equilibrium as well. For example, in configurations
without surface currents, we would still expect field lines to emerge from the core 
pointing largely along the inner boundary of the normal shell, with a field strength 
there $\sim H_b$. Consequently, we would still expect the emerging field to have a field
strength similar to what is found in equilibrium.

Our calculations have employed a very simple equation of state: the $N=1$ polytrope. Within
the context of this equation of state, if we stipulate that $\rho_b$ is fixed by microphysics
of the phase change from core fluid with free protons to a crystalline crust with protons
confined to nuclei, then $H_b$ will be independent of the stellar mass. However, since
$R$ is fixed for a (Newtonian) $N=1$ polytrope the central density is $\propto M$ and
the fractional thickness of the normal shell is $\propto 1/M$. Thus, we would expect
some variation of the surface magnetic field with stellar mass, but this would only
lead to a rather small range of values differing by less than a factor of two.

More realistically, with $H_b$ fixed the surface field strength will still be
$\propto H_b\epsb$, but the value of $\epsb$ and the proportionality constant would 
depend on details of the equation of state. We expect this to remain so even if the 
Type II region does not encompass the entire stellar core, as we have assumed here: Type 
II superconductor with an inner boundary would have different currents but the lower
boundary condition on the normal shell is sensitive primarily to $H_b$.
Nevertheless, very generally we expect that
the ``most dipolar'' case will result in field strengths $\sim 10^{12}$ G with little
variation. This is contrary to observations, which show that even the dipolar field
strengths, as indicated by the $P-\dot P$ diagram for radiopulsars (e.g.
\citealt{2012hpa..book.....L}),
vary over several orders of magnitude.

The obvious solution to this problem is that the magnetic field configurations for
many pulsars are very similar to the most dipolar configuration, but that there may
be substantial deviations. We envision several reasons for such variation:
\begin{enumerate}
\item The strong Type II limit fails because $B\sim H$ in the core. We expect that this
is true for magnetars, for example.
\item Failure of the strong Type II limit is an extreme example of a case that is outside
the scope of the calculations in this paper. There are configurations with
$B\ll H_b$ that violate our requirements: (a) If $\epsb^{-1}\gg\beta\gtrsim 1$
field lines impinging from the core affect the surface field non-perturbatively. 
(b) If $\beta$ is small enough that $\ell/R\gtrsim B/H_b\sim\beta\epsb$ then 
field lines curve significantly within the transition region between the Type II
core and normal shell, which would affect their coupling.
\item Toroidal magnetic fields, which were not considered here, can alter the 
poloidal field configuration even in hydrostatic equilibrium. We will treat this
possibility elsewhere. Toroidal fields also mitigate instability. We have seen that there
is a region in the normal shell that would be unstable if the normal shell were a pure fluid
(see Fig. \ref{fig:Field-finiteB}) although in a crystalline crust shear stresses should suffice to ensure
stability there. However, the Muzikar-Pethick-Roberts instability 
 in the Type II core
\citep{PhysRevB.24.2533,ROBERTS01081981,2008MNRAS.383.1551A}
might require toroidal fields to ensure stability.
\item Even if none of the last three caveats holds, so the core is in the strong Type II
regime and toroidal fields do not affect the poloidal fields significantly, the requirement
that field lines hit the outer boundary of the Type II core vertically that is the basis for the
most dipolar configuration is very special. Relaxing it could lead to greater diversity
in surface field strengths.
\end{enumerate}
We expect that of these options the last is most important for neutron stars with surface magnetic
field strengths substantially below $10^{12}$ G. The extreme case is when field lines hit
the outer boundary of the Type II core radially. The field line configuration within the core
is shown in Fig. \ref{fig:Field-radial}, Panel a. In this extreme, the field remains radial
as it enters the normal shell, and is weak in strength, $\sim B\ll H_b$. Entering
field lines carry current density $\sim c\rho_bH(0)/\rho(0)R$ with them, and therefore 
curve downward toward the stellar equator within a short distance of the boundary:
crudely the field should penetrate a distance $\sim B\rho(0)R/\rho_bH(0)$, which is
$\sim\beta(\rho_b/\rho(0))^{b-1}\ll 1$ times the thickness of the shell.
We might then expect that the entering field lines are confined to a small region of the
normal shell, and the field strength at the surface could be $\lesssim B\epsb$, which may
be very small. We will report elsewhere on calculations that employ different boundary conditions
at the outer radius of the Type II core.

Even though it relies on special conditions, the ``most dipolar'' solution reveals that there
is a characteristic surface magnetic fields strength that is determined by the nuclear
equation of state. Our calculations show that this field strength is $\sim 10^{12}$ G, 
reassuringly close to the values deduced for many neutron stars. By relaxing the restrictive
boundary conditions underlying the most dipolar solution, it seems plausible that 
a variety of field strengths may be attained in equilibrium. Field strengths $\gg 10^{12}$ G
or $\ll 10^{12}$ G may be attained under other restrictive conditions.

\section*{Acknowledgements}
We thank S. Lander and A. Sedrakian for helpful correspondence.
This research was supported in part by NSF grant AST-0606710, NSF grant
PHY11-25915, by a fellowship from
the NASA/New York Space Grant Consortium and by the College of Arts and Sciences,
Cornell University.

\bibliographystyle{mn2e}
\bibliography{equilibrium}   

\appendix

\section{Perturbation Solutions}
\label{aperturbsoln}

We begin with a zeroth order solution, $A_0(x,\theta)$, that corresponds to
a given $\Jchshell(A)$; we use \eq{mostdipolar}.
Perturbations distort the solution to
\be
A(x,\theta)=A_0(x,\theta)+\sin\theta\sum_{J=0}^\infty\dA_j(x)\pj(\cos\theta)~.
\ee
The perturbed solution corresponds to $\Jch(A)=\Jchshell+\dJch(A)$, where
$\dJch(A)$ must be determined along with the perturbed field solution.
To do this, we begin by choosing a functional form for $\dJch(A)$ which
will necessarily involve unknown parameters to be determined. Since
$A\approx A_0$ by assumption, we substitute $\dJch(A)\approx\dJch(A_0)$
to find the perturbed field; because $A_0$ is known we may expand
\be
\sin\theta\dJch(A_0)=\sum_{j=0}^\infty S_j(x)\pj(\cos\theta)
\label{dJchAz}
\ee
to get the equation
\be
{d^2\dA_j\over dx^2}-{(2j+1)(2j+2)\dA_j\over x^2}=-{H(0)R^2 x\sin x S_j(x)
\over\pi^2}~,
\ee
which, with the boundary condition 
$xd\dA_j/dx+(2j+1)\dA_j=0$, has the solution
\ba
\dA_j(x)&=&{\dA_j(x_b)x_b^{2j+1}\over x^{2j+1}}+
{H(0)R^2\over (4j+3)\pi^2}\Biggl[x^{2j+2}\int_x^\pi{dx'\sin x'
S_j(x')\over(x')^{2j}}
\nonumber\\& &
-{x_b^{4j+3}\over x^{2j+1}}\int_{x_b}^\pi
{dx'\sin x'S_j(x')\over(x')^{2j}}
+{1\over x^{2j+1}}\int_{x_b}^x dx'(x')^{2j+3}\sin x'S_j(x')\Biggr]~.
\label{perturbedsoln}
\ea
Differentiating Eq. (\ref{perturbedsoln}) and evaluating at $x_b$
implies
\be
-{R^2\dH_{\theta,j}(x_b)\over\pi^2}
=\left({1\over x}{d\dA_j\over dx}\right)_{x_b}
=-{(2j+1)\dA_j(x_b)\over x_b^2}+
{H(0)R^2x_b^{2j}\over\pi^2}
\int_{x_b}^\pi {dx'\sin x'S_j(x')\over(x')^{2j}}
\label{Hatxb}
\ee
where the perturbed field tangential to the boundary at $x_b$ is
\be
\dH_\theta(x_b,\theta)=\sum_{j=0}^\infty \dH_{\theta,j}(x_b)\pj(\cos\theta)~.
\label{dHatxb}
\ee
Eq. (\ref{Hatxb}), given Eq. (\ref{dHatxb}), is used to determine the
unknown parameters of the source term, $\dJch(A)$. Once these parameters
have been found, Eq. (\ref{perturbedsoln}) evaluated at $x=\pi$
determines the multipole moments $Q_j=\dA_j(\pi)R^{2j+1}$:
\be
Q_j
=\dA_j(x_b)r_b^{2j+1}+
{H(0)r_b^{2j+3}\over 4j+3}
\int_{x_b}^\pi dx'\sin x' S_j(x')\left[\left({x'\over x_b}\right)^{2j+3}
-\left({x_b\over x'}\right)^{2j}\right]~.
\label{multipoles}
\ee
Non-trivial perturbations will always engender multipole fields at the
surface.

\subsection{Surface Currents from Transition Layer}
\label{transitionfield}

We write the zeroth order solution as $A_0=A_s\ftil(x;x_b)\sin^2\theta$,
where $A_s=F_0(\pi)$ so $\ftil(\pi;x_b)=1$,  and
expand
\be
\dJch(A)=\sum_{j=0}^\jmax \dJch_n\left({A\over A_s}\right)^n
\approx\sum_{J=0}^\jmax\dJch_n(\sin\theta)^{2n}[\ftil(x;x_b)]^n
\label{dJchexpand}
\ee
where we substituted $A\approx A_0$; then
\ba
\Scal_j(x)&=&\sum_{n=j}^\jmax {N_{nj}\dJch_n[\ftil(x;x_b)]^n\over N_j}~,
\nonumber\\
N_{nj}&=&{2\pi\Gamma(n+2)\Gamma(n+1)
\over\Gamma(n+j+{5\over 2})
\Gamma(n-j+1)\Gamma(j+1)\Gamma(-j-{1\over 2})}
\label{Sj}
\ea
For sufficiently small transition region thickness $\ell$ the jump condition
from Type II core to normal shell is given by Eq. (\ref{jump}), and therefore
\be
\dH(x_b,\theta)={\pi\ell\Jch_b(\theta)H(0)\sin x_b\sin\theta\over R}
\ee
where $\Jch_b(\theta)$ is the current density parameter at the outer edge
of the Type II core, which we compute for each field line; then with
\be
\sin\theta\Jch_b(\theta)=\sum_{j=0}^\infty\Jch_{b,j}\pj(\cos\theta)
\ee
Eq. (\ref{Hatxb}) becomes
\be
-{\pi\ell\Jch_{b,j}\sin x_b\over R}+\frac{(2j+1)\dA_j(x_b)}{H(0)r_b^2}=
\sum_{n=j}^\jmax {N_{nj}\dJch_n\over N_j}\int_{x_b}^\pi{dx\sin x
[\ftil(x;x_b)]^n\over(x/x_b)^{2j}}~,
\label{findJn}
\ee
which is a linear equation for the $\dJch_n$. The multipole moments are
then found by combining Eqs. (\ref{multipoles}), (\ref{Sj}) and the
solution of Eq. (\ref{findJn}).

\subsection{Field Lines Poking in from the Core}
\label{bpokesin}

We do not know $A(x_b,\theta)$ {\sl a priori}. In the strong
Type II core, field lines are labelled by their foot points
$x_0$, and our solutions determine (i) the value of $\theta$
where they hit the boundary and (ii) the value of
$\Jch_b(\theta)$ there, subject to whatever constraint
we impose on the solution, such as for the ``most dipolar''
solution. We need $A(x_b,\theta)$ in order to map
$\Jch_b(\theta)\rightarrow\Jchcore(A)$, which is needed to
determine the effect of impinging field lines
on the normal shell solution.
%
%
%

\subsubsection{$A(x_b,\theta)=\onehalf Br_b^2\sin^2\theta$}
\label{dipolepokesin}

%
In this case, the background solution is \eq{AzB} with current
parameter \eq{Jchshellgen}. We take $\dH_j(x_b)=0=\dA_j(x_b)$.

%
A key feature of this background solution is that field lines 
from the Type II core penetrate into the normal shell, bringing
along the currents $\Jchc(A)=\Jch_b(\theta(A))$ 
associated with them, where $\theta(A)=\sin^{-1}(\sqrt{2A/Br_b^2})$.  
Defining $\dJchc(A)=\Jchc(A)-\Jchshell(B)$ we have
\ba
\dJch(A)&=&\dJchc(A)\Theta(\onehalf Br_b^2-A)+\dJchs(A)
\Theta(A-\onehalf Br_b^2)
\nonumber\\
&\approx&\dJch(A_0)\Theta(\thm-\theta)
+\dJchs(A_0)\Theta(\theta-\thm)~,
\label{deltaJ}
\ea
where the second line assumes $A\approx A_0$, and
\ba
S_j(x)={2\over N_j}\Biggl[\int_{\cos\thm}^1 d\mu\sqrt{1-\mu^2}\dJchc(A_0)
\pj(\mu)
\nonumber\\
+\int_0^{\cos\thm} d\mu\sqrt{1-\mu^2}\dJchs(A_0)\pj(\mu)\Biggr]~.
\label{Sjpokesin}
\ea
Generically, the first term in Eq. (\ref{Sjpokesin}) is
${\cal O}(1)$ for $x-x_b\lesssim\beta\epsb$ and is ${\cal O}(\beta^2)$
for $x-x_b\sim\epsb$.
To find $\dJchs(A)$
we must first specify its functional form
and then determine the associated parameters from Eq. (\ref{Hatxb})
with $\dH_{\theta,j}(x_b)=0$; the generic scalings suggest that
$\dJchs(A)={\cal O}(\beta)$. If we expand 
\be
\dJchc=\sum_{n=0}^\jmax \dJchcn\sin^{2n}\theta
\ee
on $x_b$, with {\sl known} coefficients $\{\dJchcn\}$ and
$\dJchs(A)$ as in 
Eq. (\ref{dJchexpand}) with {\sl unknown} coefficients $\{\dJchsn\}$ 
\ba
S_j(x)&=&\frac{1}{N_j}\sum_{n=0}^{\jmax}
\left\{\Dnj\dJchcn+\left[\frac{N_{nj}[\beta x_b/x+\ftil(x;x_b)]^n-\beta^n\Dnj}{(\beta r_b/R+1)^n}\right]
\dJchsn\right\}
\nonumber\\
\Dnj&\equiv&
\frac{2}{\sin^{2n}\thm}\int_{\cos\thm}^1 d\mu(1-\mu^2)^{n+1/2}\pj(\mu)~;
\label{Sjeval}
\ea
although
$N_{nj}=0$ for $j>n$, $\Dnj\neq 0$ in general.
%
The $\{\dJchsn\}$ are the solution of
\ba
0&=&
\sum_{n=0}^\jmax\left(\dJchcn-\frac{\beta^n\dJchsn}{(\beta r_b/R+1)^n}\right)\int_{x_b}^\pi 
\frac{dx\sin x\Dnj}{(x/x_b)^{2j}}
\nonumber\\
& &+\sum_{n=j}^\jmax N_{nj}\dJchsn\int_{x_b}^\pi
\frac{dx\sin x[\beta x_b/x+\ftil(x;x_b)]^n}{(x/x_b)^{2j}(\beta r_b/R+1)^n}~.
\label{dJchsneq}
\ea

\subsubsection{General $A(x_b,\theta)$}
\label{generalpokesin}

We can incorporate more complicated $A(x_b,\theta)$; let
\be
A(x_b,\theta)=\onehalf Br_b^2\left[\sin^2\theta+
\sin\theta\sum_{j=1}^\infty a_j\pj(\cos\theta)\right]~;
\label{Abgen}
\ee
the coefficients $\{a_j\}$ may be derived from a given $A(x_b,\theta)$
via inversion.
Eq. (\ref{Abgen}) implies a different mapping $\theta(A)$ on $x_b$,
hence a different form of $\Jchc(A)$ for impinging field lines.
With Eq. (\ref{Abgen}), the background solution is multipolar. The $j=0$
component is Eq. (\ref{AzB}) but the other multipoles are generically
\ba
A_j(x)&=&{Br_b^2a_jx_b^{2j+1}\over 2x^{2j+1}}+
{H(0)R^2\over (4j+3)\pi^2}\Biggl[x^{2j+2}\int_x^\pi{dx'\sin x'
\Sz_j(x')\over(x')^{2j}}
\nonumber\\& &
-{x_b^{4j+3}\over x^{2j+1}}\int_{x_b}^\pi
{dx'\sin x'\Sz_j(x')\over(x')^{2j}}
+{1\over x^{2j+1}}\int_{x_b}^x dx'(x')^{2j+3}\Sz_j(x')\Biggr]~.
\label{multzero}
\ea
The current density in this model is $\Jchshell+\Jchz(A)$, where
$\Jchz(A)$ must be determined. Because the impinging fields are
weak compared with $H_b$, we can determine the currents perturbatively
via the analogues of Eqs. (\ref{dJchAz}), (\ref{dJchexpand}) and 
(\ref{Sj}), plus the analogue of Eq. (\ref{Hatxb}) with the condition
$\dH_j(x_b)=0$.

Once the full background solution has been found, the boundary of the
region to which impinging field lines are confined may be determined.
We can then proceed as in \S\ref{dipolepokesin} to find the perturbation
to the background solution. Eq. (\ref{Abgen}) implies a total flux
$A(x_b,\pi/2)$ through the upper hemisphere. Consistency of the perturbation
theory requires $A(x_b,\pi/2)<<H_bR^2\epsb$. 

\section{The Thin Shell Approximation for the Transition Layer}
\label{transition}

The proton density plummets in a thin layer that extends from $r=r_b$ 
to $r=r_b+\ell$; let $r=r_b+\ell s$. 
As $n_p$ falls, so does $\Delta_p$ and the superconductor disappears.
For small enough
$\pfp$, nonzero temperature will matter: for $\Delta_p
\sim T$, the superconductor may first become Type I before
disappearing entirely at $\pfp\neq 0$. Our treatment here
assumes that $T\neq 0$ is only important in a very thin subdomain
of the transition layer that has little effect on the magnetic field.
With $\Delta_p\propto\pfp^2$ at small $\pfp$ for $T=0$
\citep[e.g.]{2005NuPhA.763..212A}
\eq{typeiicond} implies
$\kappa\propto\pfp^{-1/2}$, so the superconductor remains Type II.
However, as $k_L$ shrinks interactions among field lines become 
increasingly important; if $\Hc(\rho)=\Hc(s)$ within the layer,
then the magnetic field strength is
\be
H=\Hc(s)+B\fHB
\label{Heqn}
\ee
where $\Hc(s)$ is given by \eq{Hval} and $f(0)=1$ and $f(z)\to0$
exponentially as $z\to\infty$.
%
%
Amp\'ere's law is
\be
\frac{\pi\ell\sin x_b H(0)\Jch(A)\sin\theta}{R}=\frac{\partial H_\theta}
{\partial s}+\frac{\ell}{r}\left(H_\theta-
\frac{\partial H_r}{\partial\theta}\right)~;
\label{ampthinlayer}
\ee
if we assume that field lines do not rotate significantly, so that we
can identify $A$ with $\theta$, and only retain the lowest order in
$\ell$, \eq{ampthinlayer} implies
\be
-H_\theta(s,\theta)\simeq
\left[H_b-\frac{\pi\ell s\sin x_b H(0)\Jch(A)}{R}\right]\sin\theta~.
\label{Hthetatrans}
\ee
The magnetic induction field is
\be
\Bvec=-\frac{\thetahat}{\ell r\sin\theta}\frac{\partial A}{\partial s}
+\frac{\rhat}{r^2\sin\theta}\frac{\partial A}{\partial\theta}~;
\label{Bthinlayer}
\ee
in the thin layer approximation we assume that $B_r$ remains constant
and therefore $B_\theta=\pm\sqrt{1-B_r^2/B^2}$. We assume that 
$H_\theta$ (and $B_\theta$) remain in the $-\thetahat$ direction, so 
\be
\left[H_b-\frac{\pi\ell s\sin x_b H(0)\Jch(A)}{R}\right]\sin\theta
\simeq\sqrt{1-\frac{B_r^2}{B^2}}\left[\Hc(s)+B\fHB\right]~,
\label{solvefordAds}
\ee
which is an algebraic equation for $B$. Once $B$ is found from 
\eq{solvefordAds}, we can solve for the change in $A$:
\be
\frac{\partial A}{\partial s}\simeq\ell r_b\sin\theta\sqrt{B^2-B_r^2}~.
\label{dAdseqn}
\ee
Ultimately, $B$ approaches \eq{Hthetatrans}, hence the change in $A(s,\theta)$
across the layer is
\be
\Delta A(\theta)=A(1,\theta)-A(0,\theta)\simeq \ell r_b\sin^2\theta\left[H_b
-\frac{\pi\ell\sin x_b H(0)\Jch(A)}{2R}\right]~.
\label{changeinAlayer}
\ee
\eq{changeinAlayer} shows that $\Delta A(\theta)$ is relatively 
small provided that $\ell H_b/B r_b\ll 1$.


\label{lastpage}
\end{document}